# Online monitoring of local taxi travel momentum and congestion effects using projections of taxi GPS-based vector fields

**Xintao Liu**
*PhD, Assistant Professor*
*Department of Land Surveying and Geo-Informatics, The Hong Kong Polytechnic University, Kowloon, Hong Kong*
*xintao.liu@polyu.edu.hk*

**Joseph Y.J. Chow**
*PhD, Assistant Professor & Deputy Director*
*C2SMART University Transportation Center*
*Department of Civil & Urban Engineering, New York University, New York, NY, USA.*
*Corresponding author: joseph.chow@nyu.edu*

**Songnian Li**
*PhD, Professor*
*Department of Civil Engineering, Ryerson University, Toronto, ON, Canada*
*snli@ryerson.ca*

**Acknowledgements**
The authors wish to thank the DATATANG Company for the real-time GPS data used in this study. Dr. Xintao Liu acknowledges the funding support from an Area of Excellence project (1-ZE24) and a startup project (1-ZE6P). Dr. Chow is partially supported by the C$^2$SMART Tier 1 University Transportation Center, which is gratefully acknowledged.

**The Conflict of Interest**
The authors declare that there is no conflict of interest regarding the publication of this paper.

*Accepted for publication in Journal of Geographical Systems*

## ABSTRACT

Ubiquitous taxi trajectory data has made it possible to apply it to different types of travel analysis. Of interest is the need to allow someone to monitor travel momentum and associated congestion in any location in space in real time. However, despite an abundant literature in taxi data visualization and its applicability to travel analysis, no easy method exists. To measure taxi travel momentum at a location, current methods require filtering taxi trajectories that stop at a location at a particular time range, which is computationally expensive. We propose an alternative, computationally cheaper way based on pre-processing vector fields from the trajectories. Algorithms are formalized for generating vector kernel density to estimate a travel-model-free vector field-based representation of travel momentum in an urban space. The algorithms are shared online as an open source GIS 3D extension called VectorKD. Using 17 million daily taxi GPS points within Beijing over a four-day period, we demonstrate how to generate in real time a series of projections from a continuously updated vector field of taxi travel momentum to query a point of interest anywhere in a city, such as the CBD or the airport. This method allows a policy-maker to automatically identify temporal net influxes of travel demand to a location. The proposed methodology is shown to be over twenty times faster than a conventional selection query of trajectories. We also demonstrate, using taxi data entering the Beijing Capital International Airport and the CBD, how we can quantify in nearly real time the occurrence and magnitude of inbound or outbound queueing and congestion periods due to taxis cruising or waiting for passengers, all without having to fit any mathematical queueing model to the data.

## 1. Background

Increasing urbanization and population growth contribute to increasingly negative externalities in cities: congestion, pollution, and vulnerability of the population and economy to extreme events. The rise of Big Data (Tansley and Tolle, 2009), especially human mobility data at the city-scale, presents a potential solution to cope with these problems. Having access to real-time locations of people throughout the city circumvents the need to connect traffic patterns to travel (and economic) patterns, and instead directly connects a real-time data source to the "pulse" of the economy. This causes a paradigm shift that presents an opportunity in how we visualize and quantify travel demand patterns through space-time.

One such popular data source is from taxi GPS trajectories (see Yue et al., 2014, for a comprehensive survey). For planning purposes, taxi GPS data can be used to mine origin-destination patterns, time of day variations for specific locations, hotspots, trip distances, etc. (Yue et al., 2009; Zheng et al., 2011; Huang et al., 2015; Tang et al., 2015; Mao et al., 2016; Shen et al., 2017); to cluster land use types (Liu et al., 2015); and to identify critical locations or quantify resilience (Zhou et al., 2015; Zhu et al., 2017). These objectives have been addressed with other similar data sources, such as call detail records (CDRs) (Toole et al., 2015).

In all the above studies cited, the use of the taxi trajectory data is for planning purposes only. They are not designed for online operations such as city monitoring (Chow, 2016), e.g. quantifying the effects of major incidents or weather in real time as they occur and querying spatial-temporal congestion patterns to identify bottlenecks to conduct remedial operational strategies. For example, if a user wanted to query all the arrivals and departures from a particular point of interest (POI) during a specific time interval, it would have to run a search through every taxi trajectory to filter out the ones that pass through that space-time perimeter. This is extremely computationally expensive, and thus can only work for planning purposes. Nonetheless, city monitoring is necessary for a smart city to thrive and taxi data can help fulfill that role (Sayarshad and Chow, 2015). No example is more evident of this need than the whole theory of macroscopic/network fundamental diagrams. This theory has arisen significantly in the last few years because of the importance of monitoring congestion at a city without having to pay specific attention to specific road facilities (e.g. Geroliminis and Daganzo, 2008; Daganzo and Geroliminis, 2008; Mahmassani and Saberi, 2013) due to the availability of large scale urban data.

However, network fundamental diagrams depend on monitoring vehicles entering and exiting a pre-defined boundary and does not consider spatial proximity. Another example of such an effort to measure dynamic activity changes using taxi trajectories is that of Scholz and Lu (2014), although their work is limited to only hotspot changes and in terms of offline analysis. An online methodology is needed to allow a user to monitor traffic congestion considering spatial proximity and query metrics for any local POI of interest.

One relevant area of research is the vectorization of taxi trajectories. Liu et al. (2012) proposed doing so for taxi trips within a time geographic (Hägerstrand, 1970) context so that a distance decay distribution law could be derived. They found that a Lévy flight distribution can be used to model these trips using data from Shanghai, China. Vectorization is potentially an effective method because directionality of trip demand can be stored at a unit interval, and then very fast computations can be run using vector calculus (integrals, summations, dot products to project vectors onto other directions, etc.).

The use of vector fields in urban setting began in the early 1970s, through the work of Angel and Hyman (1970, 1976) to construct "velocity fields". Puu and Beckmann (1999) envisioned time-space as a continuous space. Miller and Bridwell (2009) proposed a more generalized vector field theory to characterize time-geographic urban space, which is a kind of vector GIS method that considers both scalar density and directional cost functions—a field theory for individuals. Liu et al. (2014) extended the field theory to a population, as a "momentum" of travel demand, leading to a momentum vector field. In physics, momentum is the product of mass and velocity. For travel demand momentum, the "mass" is the population while the "velocity" is the travel velocity within a space-time domain. The authors estimated this vector field using 3D kernel density, called vector kernel densities. A primary advantage of such an approach is that all the inherent directionality in the demand, typically stored in high dimensional OD matrices, is now preprocessed into unit vectors in a time-space field. The result is that a query (for operational purposes) for travel momentum toward a particular POI in time, for example, can simply be computed by running a projection of the vector field toward that location in time-space. There is no need to query individual trajectories to see if they fall within some thresholds, marking this study as the first to allow GPS trajectory data for spatial-temporal monitoring for operational purposes. As an analogy, the construction of a vector field (and its subsequent online updating) is similar to storing in memory the demand patterns of the system of interest so that spatial-temporal queries can be readily extracted.

While Liu et al. (2014) demonstrated the methodology using a proof of concept from Toronto travel survey data, it was never applied to taxi GPS data in an online framework. In the case of taxi systems, an additional unique quality is that the inbound and outbound flows for any given POI is typically conserved over a long enough period of time, since taxis don't generally park long term. As a result, being able to query taxi momentum projections online has at least two useful purposes: (1) it allows a user to monitor hotspots dynamically so that operational strategies can be implemented, and (2) in doing so it can quantify in real time the taxi queueing bottlenecks that may occur at any POI upon request. No other study, to the best of the authors' knowledge, has done this before.

We propose guidelines for how to use this method for *any* real-time GPS data set around the world. We formalize a set of algorithms for generating vector kernel density based on the methodology from Liu et al. (2014) and then develop an integrated 3D analysis GIS package called *VectorKD*. VectorKD is an open source project that is available for further extension and validation for general purposes in related urban studies. VectorKD is applied to a data set of twelve thousand Beijing taxis conducting over 17 million daily trips. This allows us to demonstrate how we can identify in real time when taxi congestion effects at a local POI may begin to occur.

The remainder of this paper is organized as follows. Section 2 provides a simple description of the study area and data sources, presents the formalized algorithms for generating vector kernel density, followed by the vector kernel density projection as a part of the experimental methodology. Section 3 illustrates the results of the experiments using real-time GPS trajectories of taxis in Beijing, China. Section 4 concludes the paper.

## 2. Methodology

The theory behind the methodology of the vector fields is discussed in Liu et al. (2014), which proposes a population-based vector field for visualizing and representing time-geographic demand momentum. The field is estimated using a vector kernel density generated from observed trajectories of a sample population. For more details, please refer to the original work by Liu et al. (2015). The key term to note is that of "travel momentum", which is a vector determined by the product of a scalar population demand and a vector travel speed. Travel momentum is represented as a vector field, which in practice involves separating discrete cells, or "taxels", for each unit of time-space, and using vector kernel densities to estimate the vector field within each taxel from the taxi GPS sample data.

*2.1. Generating the vector kernel density*
Based on the real-time GPS data, we generate vector kernel density as a continuous representation of taxi travel momentum. Pseudo codes are provided to better illustrate this method. The generation of the vector kernel density is divided into four steps in our work.

First, we decompose the continuous 3D time geographic urban space into quantitative units: 3D cubes or "taxels". We use a grid to rasterize the 2D geographic space into cells (see Fig. 1a), given a spatial resolution (e.g., 100 meters). Then we use a sizable time unit/resolution (e.g., one minute or one hour) to decompose the continuous time dimension into discrete time slots, (e.g., $t_0$ to $t_1$ in Fig. 1a).

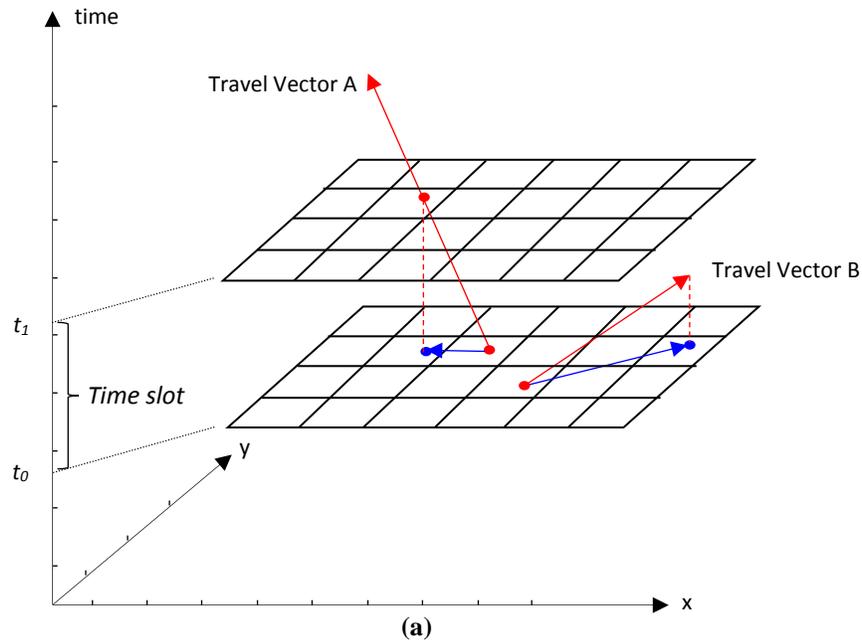

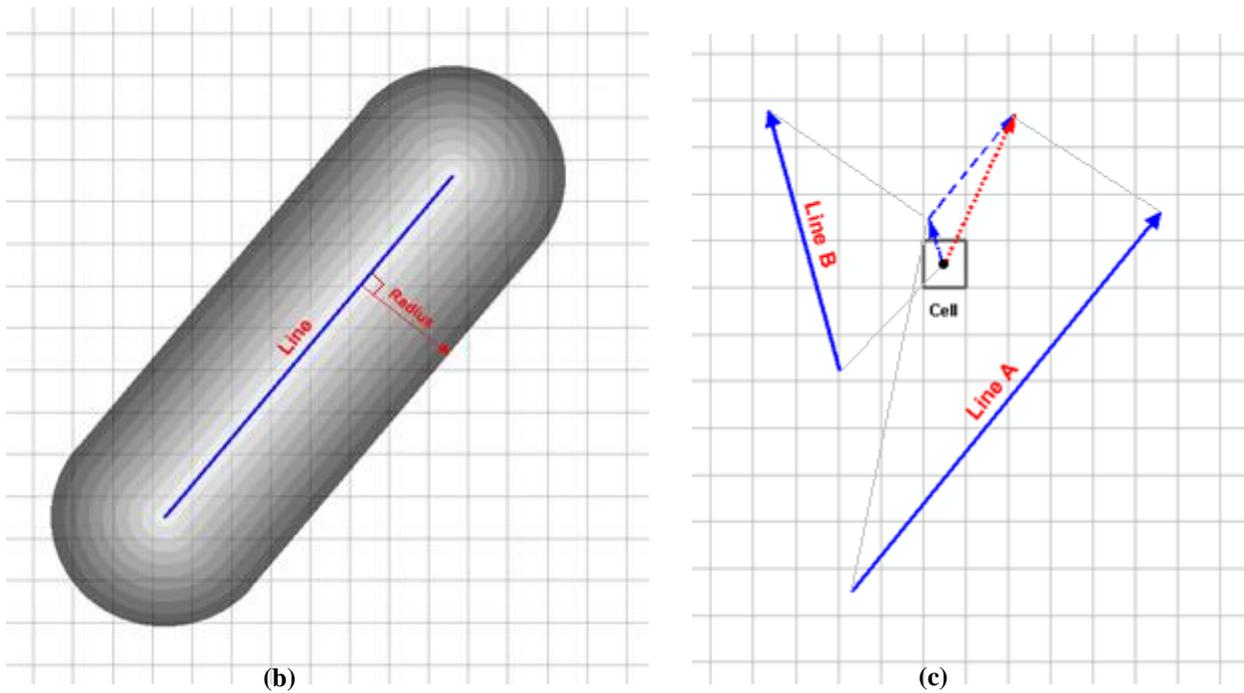

**Fig. 1**. (a) Splitting trip vectors with time slot; (b) How line based kernel density works; (c) Vector addition for vector based kernel density.

Second, we clip the travel vectors within each time slot. A travel vector (red solid lines in Fig. 1a) refers to two consecutive GPS points in a taxi trajectory. Such a travel vector has a start time as well as an end time, based on which we can select all travel vectors overlapping with a time slot in time dimension. If the time slot does not completely contain a trip vector (e.g., Trip Vector A in Fig. 1a), we clip the trip vector by time ratio. All the trip vectors within the current time slot are projected (blue solid lines in Fig. 1a) onto the 2D space.

Third, we generate a 2D kernel density map for each time slot. In Fig. 1b, it shows how the kernel density method works for 2D lines in 2D space. Conceptually, a smooth curved surface is created over each line within a given search radius (e.g., 100 or 1000 meters in Euclidean distance). The surface has the highest density value at the location of line and the value diminishes with increasing distance from the line and reaches zero at the radius distance. The closer the cells to the line, the higher the density value they have. If there are more than one line, we can simple add up all density value for each cell. In doing so, we can get a 2D kernel density estimation.

Fourth, we vectorize the 2D kernel density map. To add a vector dimension to the 2D kernel density map, we use travel vectors to generate a new vector for each cell. In Fig. 1c, it shows the conceptual process to calculate such a new vector for a cell. Suppose there are two travel vectors (blue solid lines A and B in Fig. 1c) that are within the search radius from the center point of the cell (bold square with a black dot center in Fig. 1c). For each of these two travel vectors, we can make a parallel vector that starts from the center of the cell: the short dashed blue arrow is the parallel vector of Trip Vector A and the long one is the parallel vector to Trip Vector B. Using vector addition, we can get a new added-up vector (dot red arrow). In doing so, we obtain the directional vector for the cell (i.e., the direction of the new vector). By combining the 2D density value and the cumulative vector together, we get the vector density estimation for each taxel.

The following pseudo code in Algorithm 1 is provided to summarize the above four-step method. We use $R$ as the bandwidth (or search radius) for a number of lines within a time slot and use *cell_size* to rasterize the 2D space into cells (or grid). The minimum and maximum coordinates are ($min\_x$, $min\_y$) and ($max\_x$, $max\_y$) of the 2D space and the number of travel vectors is $N$. The vector formula for the $i^{th}$ trip vector is: ($x_{i0}$, $y_{i0}$) for start point and ($x_{i1}$, $y_{i1}$) for end point.

Algorithm 1: Generate vector kernel density

*For (x = min_x; x < max_x; x+= cell_size )*
*{*
   *For (y = min_y; y < max_y; y+= cell_size)*
   *{*
 *Cx = x + cell_siz / 2.0; // center x of current cell*
 *Cy = y + cell_siz / 2.0; // center y of current cell*
 *For (i = 1; i<= N; i++)*
   *{*
    *r = distance from current cell center (Cx, Cy) to current trip line;*
   *If (r <= R)*
   *{*
     *KDE = 21.75 / ( π * R$^2$ ) * (max (0, 1 - r$^2$ / R$^2$))$^2$; // Quartic kernel density*
     *C(i, j) += KDE;   //Add the KD value to cell kernel density C(i, j)*
     *NV = [(x$_{i1}$ - x$_{i0}$)/r , (y$_{i1}$ – y$_{i0}$)/r]; // normalized parallel vector of current trip line*
     *CV(i, j) = [Cx+ (x$_{i1}$ - x$_{i0}$)/r , Cy + (y$_{i1}$ – y$_{i0}$)/r]; //Add NV to cell vector CV(i, j)*
    *}*
   *}*
 *}*
*}*

There will be m cells along x and n cells along y, and number of travel vectors N, which is m*n*N in total. And these processes will repeat for each time slice. It should be noted that: (1) the quadratic Kernel Density Estimation (KDE) function in the pseudo code can be changed to any other function such as Gaussian or uniform ones; and (2) the generated results can be saved either in self-defined data formats (e.g., XML), in an existing GIS data format (e.g., ESRI geodatabase), or a combination of the two.

The vector kernel density in a taxel is an estimate of the taxi travel momentum vector of the underlying transport system at that location in time-space. It includes a net direction from the population as well as a net magnitude. The higher the population demand for travel in that direction at that point in time-space, the larger the vector. The higher the speed of travel, the lower the slope along the time-axis, and therefore also the larger the size of the vector.

*2.2. Projection of vector kernel density*

The generated vector kernel density represents the taxi travel momentum in the study area. As a vector field, the demand directionalities are now stored in a much more compact data structure that does not require embedding any travel demand model to quantify spatial-temporal patterns. Every point in space over time of day has a vector density (if done in real time, they would update along the time of day axis), which means both the magnitude of the density and direction of travel demand are available to use.

One new application that we propose in this study is the projection of the vector onto any POI by request in real time. The process is shown in Fig. 2. Suppose we have a POI (red point at the center) and we need to project vector kernel density of four cells onto it. For cell A, we connect the center of cell A to the POI to form a Cell-POI vector (green solid arrow line), based on which the vector kernel density of cell A (red solid arrow line started from center of Cell A) can be projected onto the cell-POI vector as the dashed blue arrow line (also started from Cell A). Mathematically, the above process is the dot product of Cell-POI vector $V_{cp} = [A1, A2]$ and cell vector $V_c = [B1, B2]$. By repeating the above projection process, we can get all projected vectors of the other cells to the POI and then add up all projected vectors to get an accumulated vector kernel density *VKD* by using Eq. (1).

$$VKD = \sum_{i=1}^{n}(Ai1 \times Bi1 + Ai2 \times Bi2) \qquad (1)$$

where $n$ is the number of cells within search radius from the current cell, $[Ai1\ Ai2]$ is the $i^{th}$ Cell-POI vector, and $[Bi1\ Bi2]$ is the cell vector as mentioned above. Note that the search radius may be set to be a threshold defined by an appropriate distance decay rate if known (e.g. Liu et al., 2012).

Computationally, note that this is much faster than the conventional process of checking each existing GPS trajectory to see if and how they align with a POI, as the computation only requires vector operations.

The direction of the projected vector of a Cell to a POI can be either opposite to or towards Cell-POI vector. If the direction is opposite, it means the potential travel momentum will be decreased from this POI, e.g., Cell A and B. On the contrary, the projection of cell vector can also be towards the POI, which means the potential increased travel demand to the POI, e.g., Cell C and D. By calculating the projection vector that are away from and towards a POI separately, we can distinguish "inbound" and "outbound" travel momentum to evaluate the patterns of POI. By plotting the two values together over time, it is possible to monitor inbound and outbound travel momentum, and to identify differences. The area under between inbound and outbound projected momentum along the time axis is the numerical integral that can be calculated using Eq. (2).

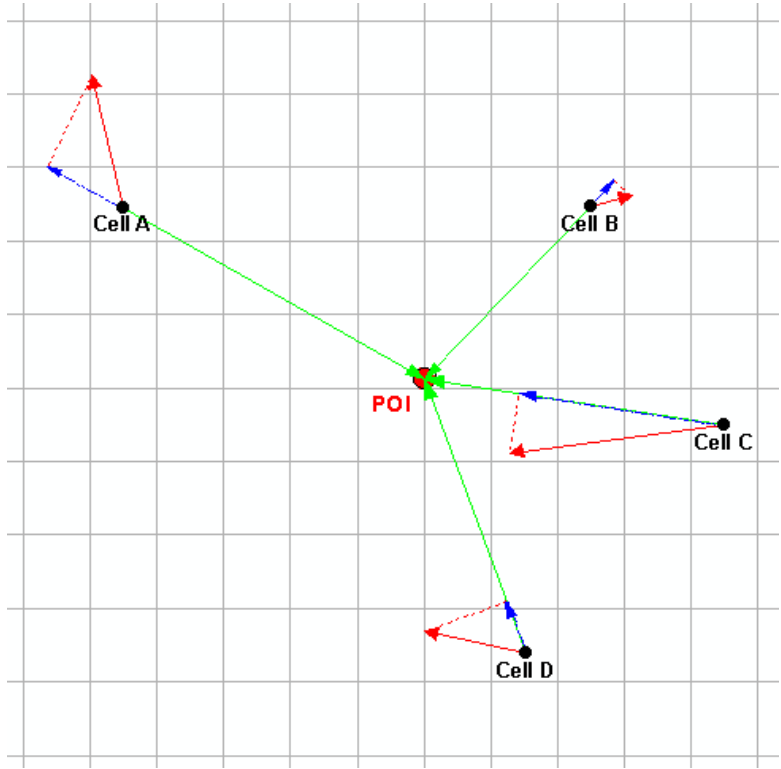

**Fig. 2**. Projection of vector kernel density onto Point of Interest (POI) as travel momentum.

$$Demand = \sum_{t=0}^{n} W_t T_t \qquad (2)$$

where $n$ is number of time slots minus one, $W_t$ is the internal width between two consecutive time slots, and $T_t$ is the average density difference of such time slots. Suppose the inbound and outbound travel momentum areas are calculated as $D_{tt}$ and $D_{ta}$. The changing rate of travel momentum between two sets of time slots can be evaluated using the following Eq. (3):

$$Rate = \frac{D_{tt} - D_{ta}}{D_{ta}} \qquad (3)$$

Note that if the outbound area becomes very small, the Rate will spike up to infinity, and when it is exactly zero the Rate would be undefined. This provides a user with a tool to automatically identify congestion events with respect to POIs and deploy different remedial strategies depending on whether the inefficiencies are caused by inbound or outbound congestion. More formally:

**Definition 1**: A **projection** of a vector kernel density of the taxi travel demand momentum **to or from a POI** measures the combined demand and efficiency to use the underlying transport system to get to or from the POI. A larger projected value indicates either high demand, high efficiency in travel speed, or both.

**Definition 2**: A **projection profile to or from a POI** plots the projection over time so that relative comparisons of travel momentum can be quantified. A projection profile can feature only inbound momentum, only outbound momentum, or both momentums overlaid together.

**Definition 3**. When inbound and outbound taxi travel momentum are overlaid in one profile, the **area between them** represents inefficiencies at the POI, such as queueing delay or congestion. An area in which the outbound momentum is larger than inbound momentum suggests there is a queue in getting into the POI. On the contrary, an area in which the inbound momentum is larger than outbound momentum suggests there is congestion leaving the POI.

Definition 3 works because taxis generally exhibit conservation of flow into and out of locations during operating hours (i.e. the number entering a zone should roughly equal the number exiting the zone given sufficient time). Therefore, differences between inbound and outbound momentum projections reflect relative differences in speed of vehicles. As a result, the areas between them reflect delays due to queueing or congestion effects leading to slowdown of the vehicles. In other words, if outbound taxi travel momentum is consistently lower than inbound taxi travel momentum, it means that there is congestion in the outbound movement, which may be due to delays in matching with customers, for example. Slower inbound suggests a queue entering the POI, which may exist in the case of airports where taxis are queueing up to pick up passengers. An illustration of the projection profile is shown in Fig. 3. The top profile shows a reduction in outbound speed compared to the inbound, meaning there is congestion leaving the area; the bottom one shows a mix of both.

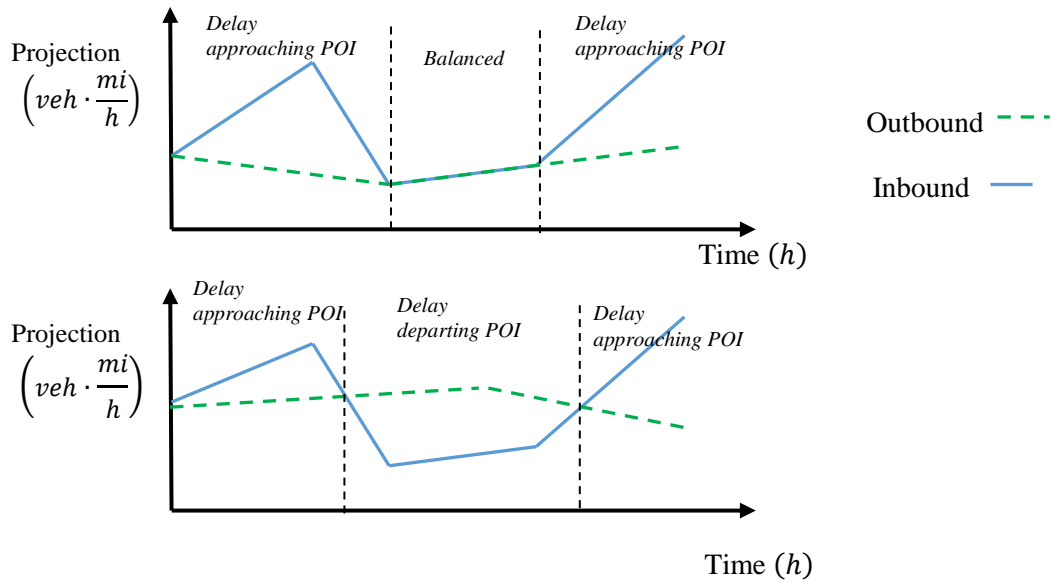

**Fig. 3**. Illustration of two different projection profiles.

## 3. Experimental design and data

We conduct several experiments to demonstrate the feasibility of generating POI-oriented patterns on-the-fly using the constructed KDE-estimated vector field. The objective is to apply the methodology in a case study and show how it can *dynamically generate queue delay for any POI* by request.

This study focuses on the built-up urban area in Beijing, the capital city of China (red star point on the left in Fig. 4) with a population of about 20 million. As shown in the middle part of Fig. 4, Beijing is encompassed by a series of ring roads. The built-up area in our study refers to the area within the outmost 6$^{th}$ ring road. The actual GPS data that was employed in this study is collected from over 12,000 taxis in Beijing during 2$^{nd}$ – 5$^{th}$, November, 2012. Each minute, the GPS enabled device installed on each taxi automatically sends its current location (i.e., latitude, longitude) and other attributes (e.g., taxi ID, timestamp, operation status, speed, direction and GPS status, etc.) to a centralized server. There are approximately $12,000 \times 24 \times 60 = 17,280,000$ GPS points per day. The trajectory of a taxi is composed of a sequence of such time-stamped GPS points in chronological order. In this study, we define any two consecutive GPS points in a trajectory as a travel vector.

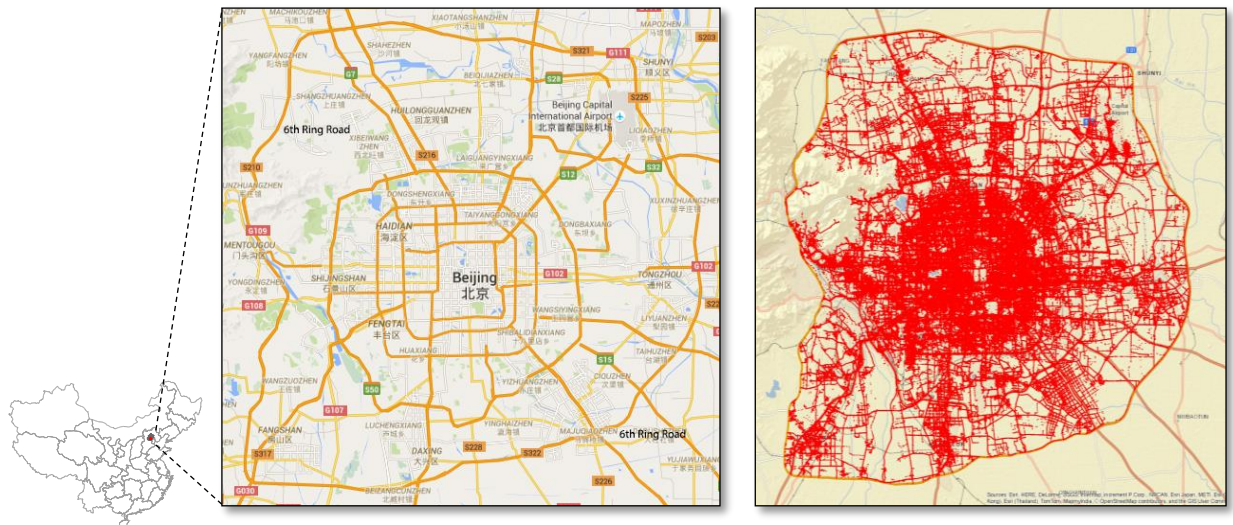

**Fig. 4**. Study area: built-up area within the 6$^{th}$ ring road in Beijing, China (left) and total daily real-time GPS locations from 12,000 taxis (right) on Nov. 2, 2012.

The sample data on Friday, November 2$^{nd}$, 2012 is shown on the right in Fig. 4. Based on the formalized algorithms described in Section 2, we design and develop an integrated 3D analysis ESRI ArcScene extension, namely VectorKD (https://github.com/xintao/3DKernel), to help generate vector kernel density and analyze travel demand patterns in Beijing. This easy-to-use ArcGIS extension tool is also useful for interested readers in relevant research, and the shared source codes are available for validation and extension.

We use the VectorKD tool to create a 200 by 200-meter grid for the Beijing study area, and choose one hour as the time resolution to create 3D taxels representing 3D time geographic space, since one hour is the basic sizable time resolution that is easy for people to understand travel demand patterns over time of day. We chose this spatiotemporal resolution because we found that was computationally more manageable for constructing the initial KDE vector field. In a mainstream-configured desktop computer, it takes around fifteen minutes to generate the initial vector field for the whole day. However, if real-time data keeps feeding into the GIS software, the vector field is locally updated in nearly real time. Then we apply the algorithms to actual GPS trajectories (DATATANG, 2015) over four consecutive days that cover both weekdays and weekend (Friday, Saturday, Sunday and Monday during November $2^{nd}$ – $5^{th}$, 2012) to generate vector kernel density for each day as the continuous field-based representation of travel demand in Beijing.

Based on the generated vector field in study area, the experiments are designed as follows. First is to visualize the travel momentum vector field for the Beijing taxi system during that timeframe. Second is to project travel demand momentum onto five selected POIs throughout the study area to show that the dynamic methodology can interpret distinct signatures for different POIs, even when they are close together. Third is to apply the methodology to the Beijing Capital International Airport to show how queue delays can be quantified dynamically in real time, something that no other taxi GPS-based methodology in the literature has yet to accomplish.

## 4. Results and discussion

*4.1 Travel momentum visualization of Beijing taxis*

Noticeably, though one-hour time interval is used to demonstrate the method in this work because one hour is easy to be recognized based on daily life experience, any other time interval can be defined and applied to this method. Hourly snapshots of the vector field, as estimated using the vector KDE, is shown in Fig. 5 for 24 hours on November $2^{nd}$ (Friday). For example, the snapshot at 00:00 represents vector field captured between 00:00 and 01:00. The height measures the magnitude of taxi density while the color refers to different directions of travel demand momentum. Based on this series of 3D visualizations, one can visually evaluate how overall taxi travel momentum patterns evolve in different hours. One can further explore local travel momentum anomalies. For example, at 5:00 and 6:00 in the morning, eastbound taxi travel momentum is dramatically increasing along a corridor, whereas in the evening it is more dispersed along all directions. In fact, the peak vector kernel density area at the upper-right part of the 6:00 subgraph is the international airport. Therefore, the physical meaning behind this anomaly is how travel momentum at 5:00 and 6:00 AM is distributed and related to the airport. As time goes to 7:00 AM, travel momentum starts to shift to central areas in Beijing. Since the generated vector kernel density maps are based on real-time GPS trajectories, they help empirically shed a deeper light on travel patterns. The visualization of vector kernel density in other days may be different from each other.

*4.2 Analysis of travel momentum projections*

To quantitatively evaluate the travel pattern at a specific location or area, we apply the projection method (refer to Section 2.2) to five selected POIs (shown in red in Fig. 6). Each point on the vector field surface is projected onto a POI and decayed by distance from the point to POI. After that, the projected vector on the POIs are added together to get the vector. These five points are transport hubs (e.g., the South Rail Station and Airport), a special historic site (e.g., the Forbidden City) and significant functional areas (e.g., IT Center and Central Business District (CBD)). These five representative locations are spatially distributed across the study area and functionally different, and thus are objectively compared in terms of their temporal relationships to inbound and outbound travel demand patterns.

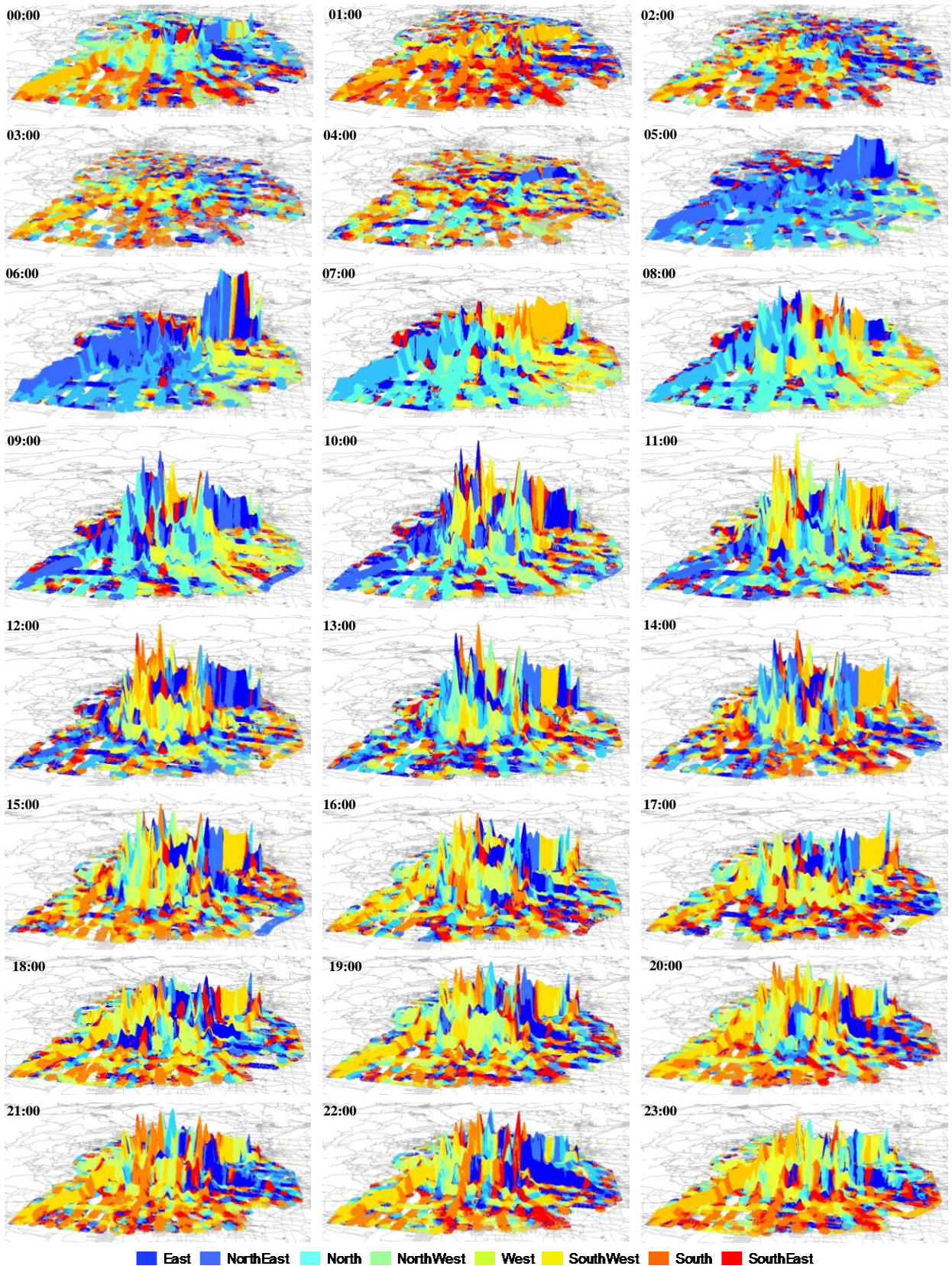

**Fig. 5**. Snapshots of vector kernel density maps in 24 hours.

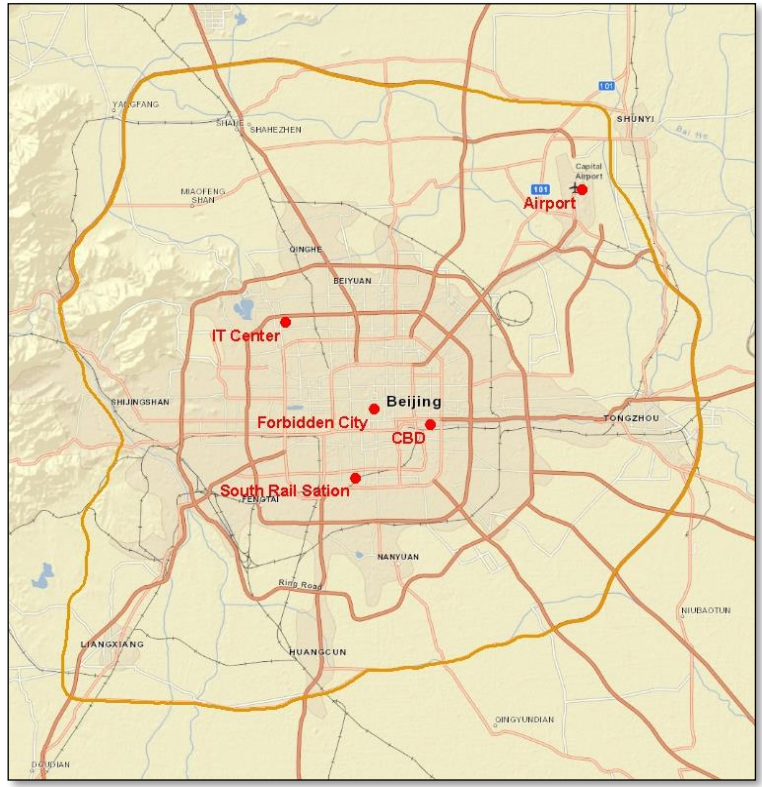

**Fig. 6**. Five selected POIs (CBD, Forbidden City, South Rail Station, Airport and IT Center).

First, we evaluate the computational efficiency of these projections. A mainstream computer configured with 64-bit Windows OS, dual processors (Intel Core i7 @ 3.6GHz) and 32GB installed memory (RAM) is used for testing. The projection for one POI takes less than three seconds, which is nearly real time. The most popular GIS platform ArcGIS 10.2 for desktop is installed in the computer to test traditional way of searching GPS trajectories. It should be noted that the GPS trajectories are organized as vectors using ArcGIS file geo-database. By comparison, the same query using traditional approach of filtering GPS trajectories would require around sixty seconds for all trajectories of 12,000 taxis. To provide a more robust computational assessment, we have now added Table 1. The computational costs of generating the initial vector field will use m cells along x and n cells along y, and number of travel vectors N, which is m*n*N in total. These processes will repeat for each time slice, and the processing time can be easily estimated."

Table 1. Comparison of computational efficiency

|  | **Number of trajectories** | **Pre-processing time** | **Operation** | **Platform** |
|---|---|---|---|---|
| Traditional approach | 12,000 | > 60s | Query | 64-bit Windows OS, dual Intel Core i7 @3.6GHz; 32GB RAM |
| Proposed method |  | <3s | Query/projection |  |

Furthermore, we compare the POI projection profiles to see if they make sense for the functions of those POIs, and if the method can dynamically distinguish between two different functional POIs located close to one another.

As discussed in Section 2.2, the projection of the taxi travel momentum onto a POI can either be towards (inbound) or away from it (outbound). As shown in the projection profiles in Fig. 7, the CBD possesses the highest projected taxi travel momentum, and the second one is the IT Center. Since these two POIs are specialized business areas, it verifies the ability of our method to recognize distinctly different POIs from the dynamically drawn projections. Note that on the weekends the momentum related to the IT Center is equivalent or worse than that for the South Rail Station, which also makes sense.

To demonstrate that projections are not purely based on geographic location, we can compare the Forbidden City with the CBD. Although the Forbidden City is geographically very close to CBD, the momentum attracted by CBD is much higher than the one attracted by the Forbidden City. On the contrary, the South Railway Station is geographically far from the IT Center, but the travel momentum they attract is quite similar to each other, especially on weekdays. Such difference and similarity may prove effectiveness of this method. Further, according to Fig. 7a, taxi travel momentum towards the Forbidden City is lower than the one away from the Forbidden City. In fact, the patterns are distinctly different on the two weekdays compared to the weekend, which fits with intuition.

Over all time periods, the Airport has the lowest projection of travel momentum, which should be due to a combination of relatively lower demand than intra-city taxi travel, the long distance for a taxi to go from downtown, and the presence of cheap and convenient public transit service to the Airport.

Generally, the demand patterns around these POIs exhibit a temporal valley around 5:00AM each the morning, and the peak is around 9:00AM in the morning, which also fits with intuition. We see that the CBD gets a spike in momentum on Friday evening, suggesting the presence of late night social activities in the downtown at that time.

The above analysis illustrates the potential applications of the dynamic methodology based on a pre-constructed vector field. We can further examine projections in terms of separated inbound and outbound patterns over time. The CBD inbound momentum in Fig. 7a fluctuates on Monday. The Forbidden City and the Airport are very stable. The fluctuations among the CBD, IT Center and South Rail Station are not synchronized, which means the travel patterns are different from each other. Such patterns reflect the people working in these areas having different schedules and thus resulting in different travel patterns.

*4.3 Model-free measuring of queue delay from taxi trajectories in real time*

In this section, we demonstrate Definition 3 by querying the projections of taxi inbound and outbound travel momentum to and from the airport as well as the CBD to quantify queue delays dynamically in real time. We compare the inbound and outbound patterns with respect to the Airport in Fig. 9a, and with respect to the CBD in Fig. 9b. We show that it is indeed possible to identify unique patterns for inbound and outbound momentum, and also from one POI to another.

Note that inbound is always equal to or higher than the outbound for the airport, while for the CBD it is primarily the opposite phenomenon. For the airport, the delay is primarily due to a slowdown of vehicles entering the airport (as opposed to exiting the airport), leading to a much higher outbound momentum than inbound momentum. This is interpreted as taxis forming queues as they enter the airport to wait for passengers to be picked up. The high outbound momentum suggests that the airport is well designed for vehicles to exit without congestion.

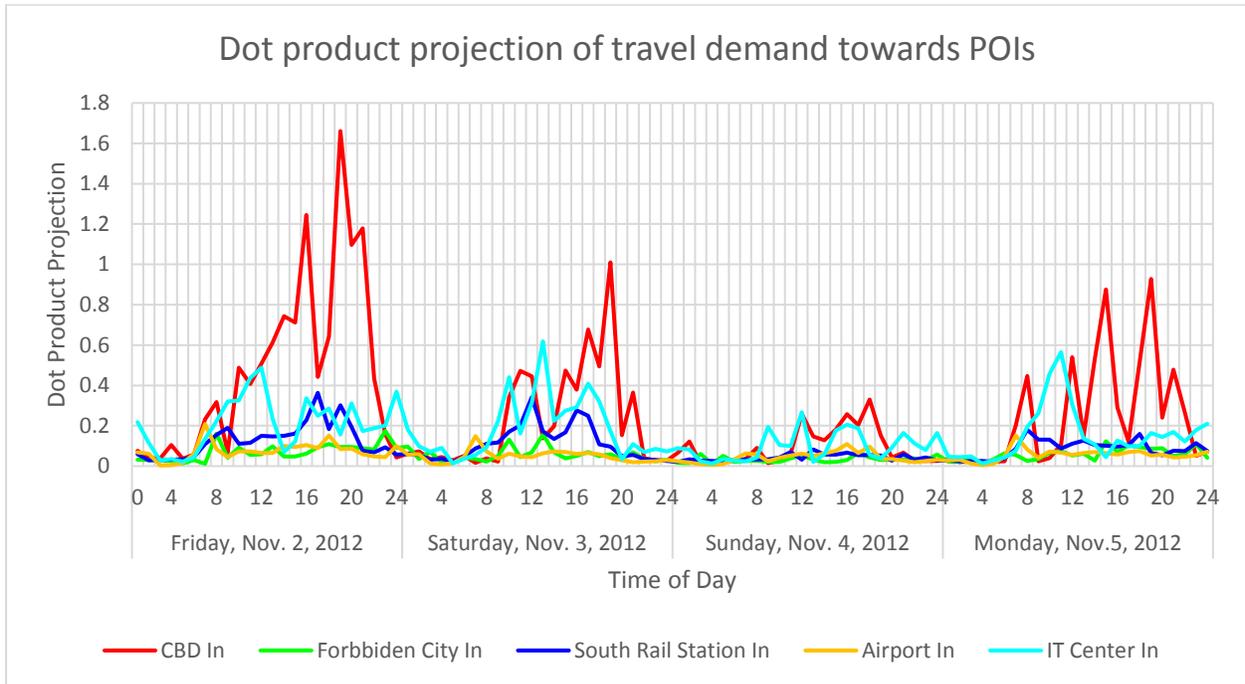

(a)

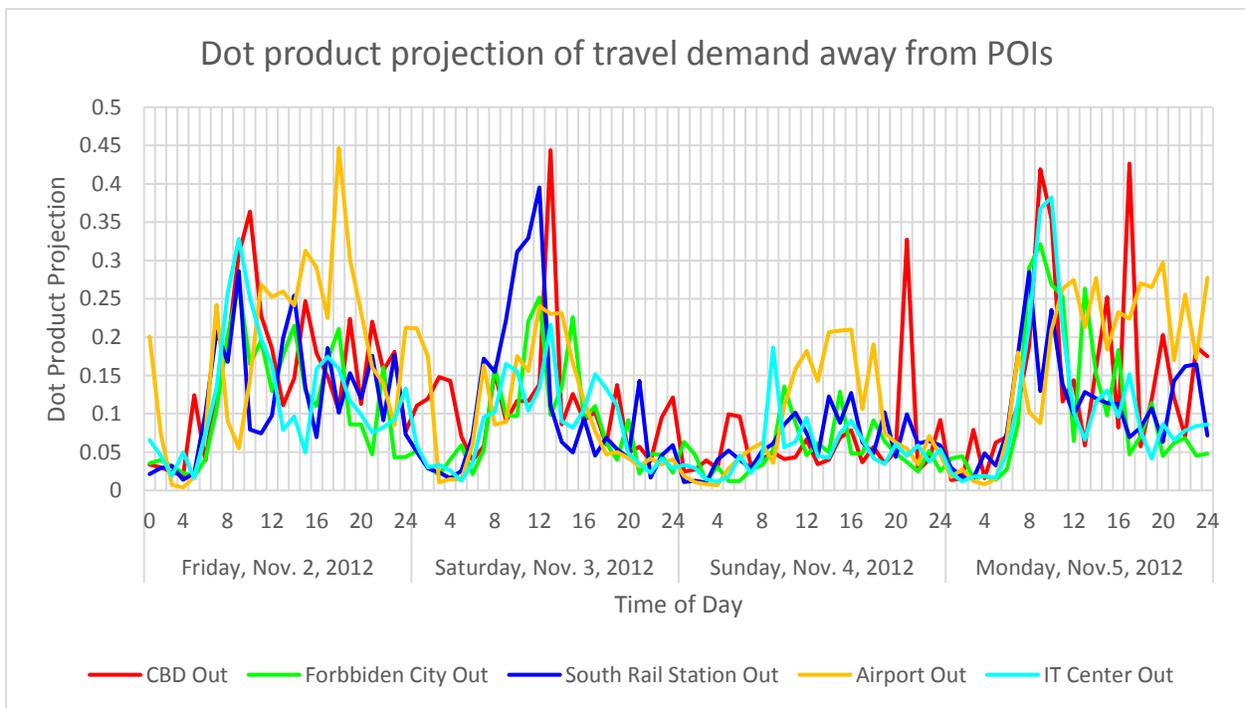

(b)

**Fig. 7**. Projection profile of taxi travel momentum onto five selected POIs in Beijing: (a) projection towards POIs; and (b) projection away from POIs.

We extract subfigures of the raw taxi GPS data and projection map in the airport area for two time slots: 8:00am and 8:00pm in Figure 8. It is obvious that looking at the data alone without projection would not be very meaningful or quantitative.

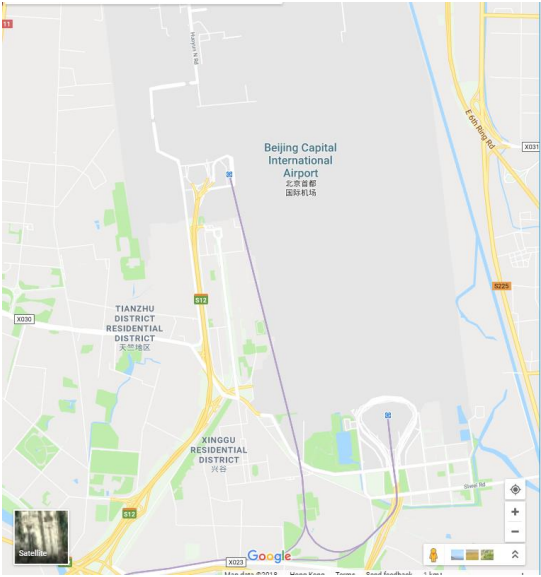

(a) The Beijing Capital International Airport area map

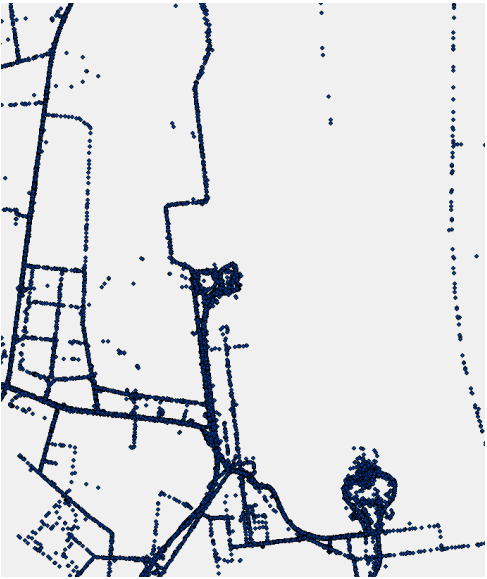

(b) Raw taxi GPS data at 8:00am

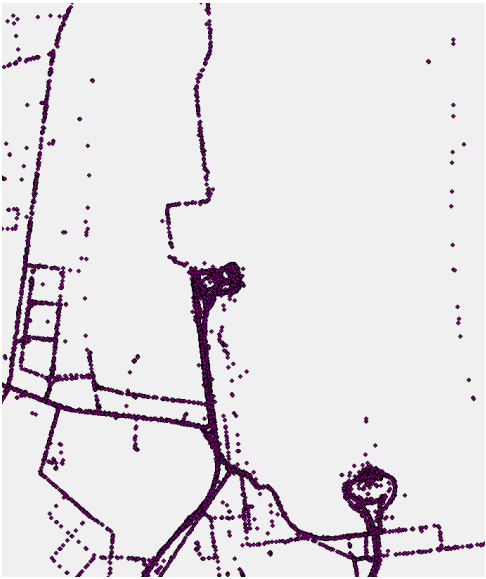

(c) Raw taxi GPS data at 8:00pm

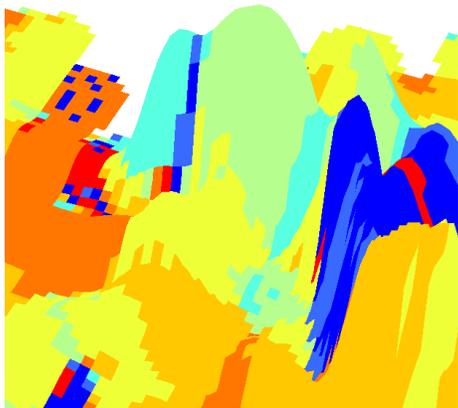
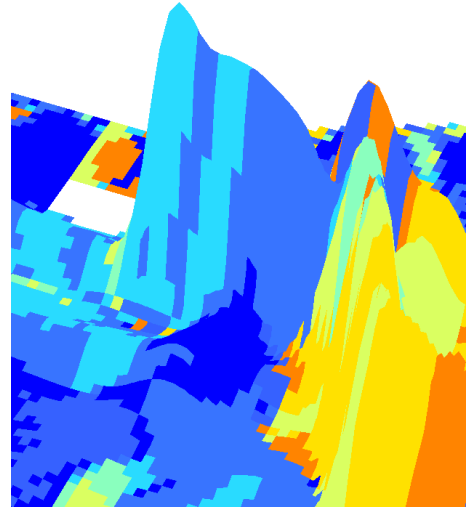

(d) 3D view of Projection map at 8:00am  (e) 3D view of projection map at 8:00pm

Figure 8. Subfigures at the airport for visualization and comparison

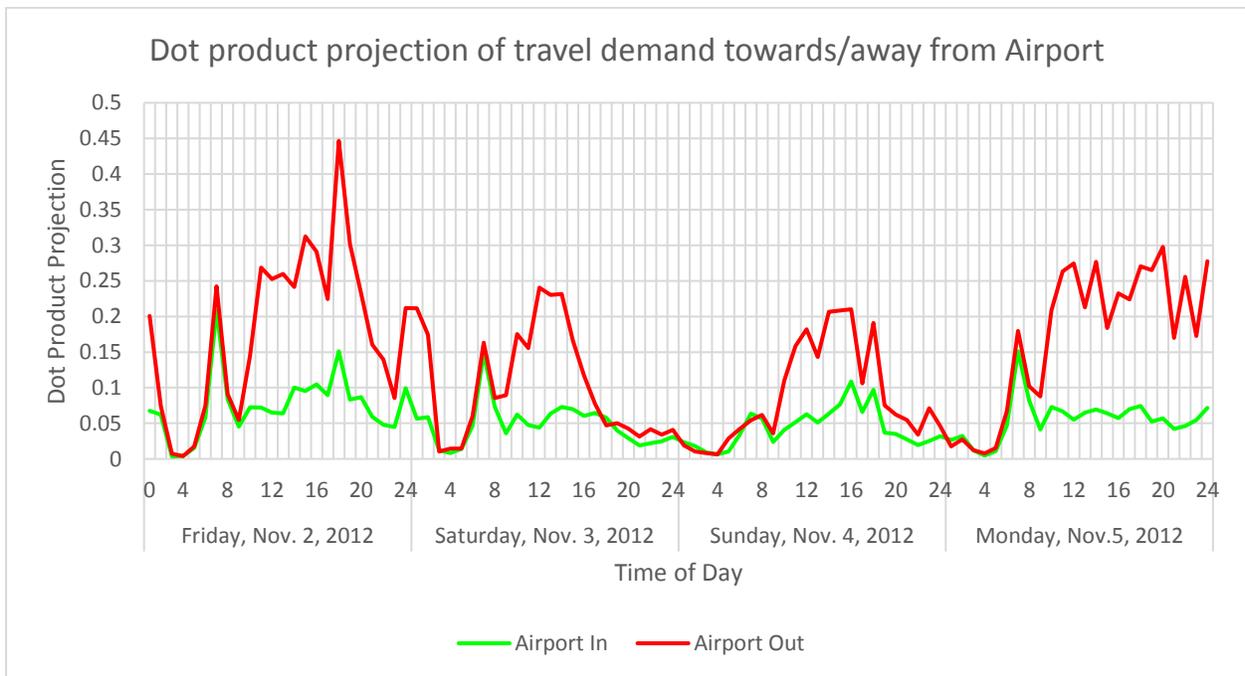

(a)

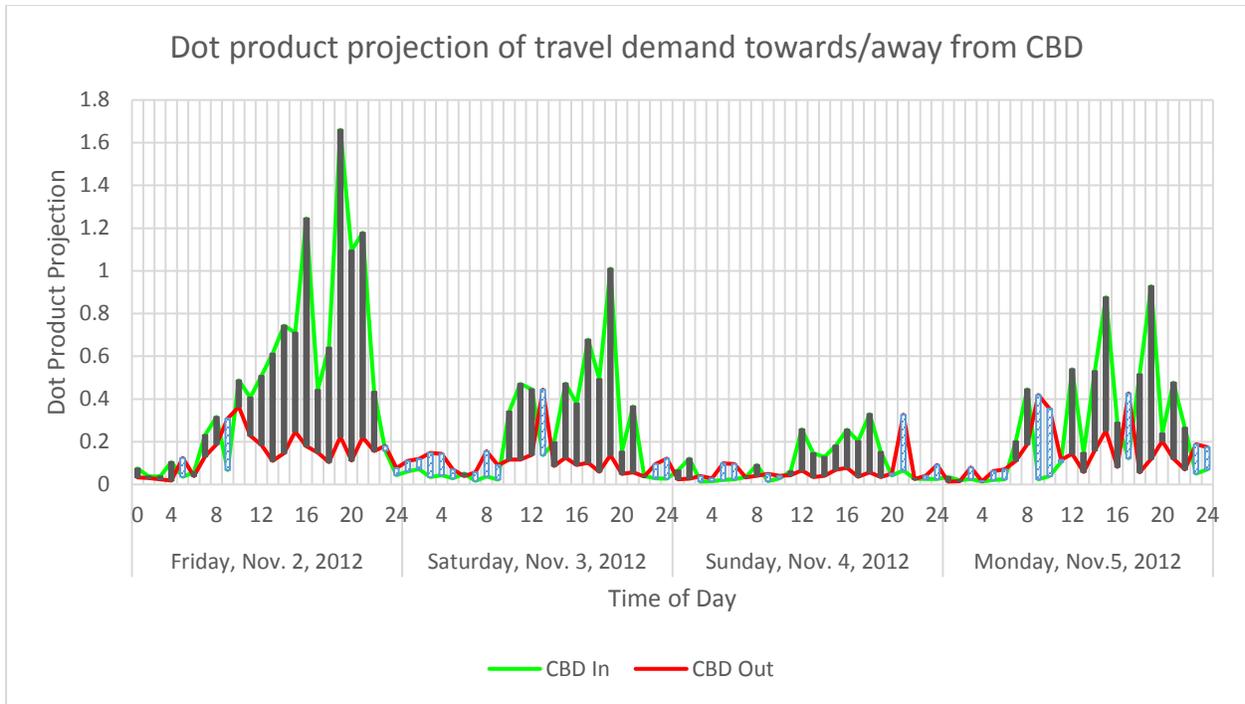

(b)

**Fig. 9**. Comparison of queue delays in projection profiles towards/away from (a) Beijing Capital International Airport, and (b) the Beijing CBD.

For the CBD, on the other hand, congestion effects occur from inefficiencies in both inbound and outbound, although outbound is much more significant. This suggests that the CBD area is highly congested during those times, resulting in bottlenecks while getting out. To try and best address this, we propose a compromise – we look at a qualitative validation by downloading the actual airline schedule for one day (see Fig. 10), because the API is only accessible for one day. I made a figure showing the number of airplane departure and arrival. The curve looks like not fit into the projected in and out travel demand, which may be because of the delay of public transit and passenger"

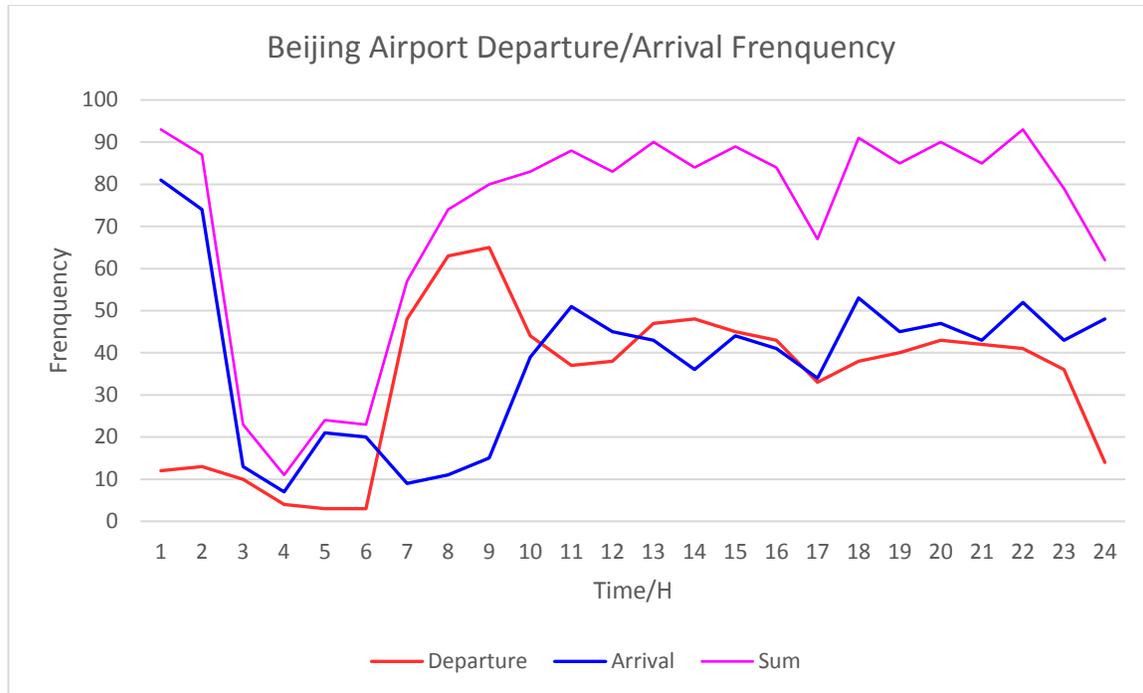

**Fig. 10**. Actual airline departure and arrival for one day at the airport as external validation of traffic demand.

This empirical analysis demonstrates that we can *dynamically quantify queueing effects and identify inbound/outbound bottlenecks* using this methodology. This is highly significant as it means that travel momentum monitoring using GPS trajectories is feasible.

## 5. Conclusion

In this study, we make several important new contributions to the literature on real-time taxi travel momentum monitoring in a time-geographic context. The most significant contribution here is the development of POI-based projection techniques using a vector field data structure for taxi GPS trajectories. The point of this methodology is to provide city agencies with a data-driven tool for monitoring that does not rely on travel models. While the method does not distinguish trip purposes of individual trajectories, it uses statistical properties to generate robust performance measures. The analogy to this method is weather stations reporting city weather conditions. Sensors are deployed at different locales; erratic patterns may result in a local site due to situational events. However, over time and using the power of large numbers, the methodology results in a reliable data feed of weather conditions for a city. In this study, we have developed a similar model-free data feed for cities for monitoring traffic conditions at different POIs based on taxi trajectory data.

The methodology is computationally efficient enough, due to preprocessing of the taxi travel momentum in the vector field, to be used in an online operational setting. The Beijing taxi case study illustrates this efficiency where queries using the vector field KDE can be obtained twenty times faster than a traditional method of searching through all the trajectories to select those that enter a POI. This sets the stage for smart cities to truly leverage on taxi GPS data to monitor cities in real time. City managers can query any specific location at any given time to see how travel momentum projects to and from that POI. Comparisons between different POIs using the Beijing taxi data verify the uniqueness of different POI projections even when located close to each other. When inbound and outbound projection profiles are overlaid together, inbound and outbound bottlenecks can be identified in real time. This allows city managers to determine the appropriate remediation strategy and schedule its deployment, such as sending in detour instructions for outbound bottlenecks or metering the inbound queues.

Many future studies can be conducted. We have demonstrated the capacity to monitor changes in real time and implemented it in an open source GIS package with real-time feeds. An online version with a real-time dashboard is to be designed for a video wall by integrating live data feeds (e.g., connected vehicles) for managing operations, which will offer researchers new insights in city monitoring. Furthermore, more urban information such as land use can be integrated to infer semantic meanings and knowledge of urban phenomena. Given the effectiveness on flow field visualizations (e.g. Parsons et al., 2013), we will also explore the potential applicability of those visualizations in future research.

**Acknowledgements**
The authors wish to thank the DATATANG Company for the real-time GPS data used in this study. Dr. Xintao Liu acknowledges the funding support from an Area of Excellence project (1-ZE24) and a startup project (1-ZE6P). Dr. Chow is partially supported by the $C^2$SMART Tier 1 University Transportation Center, which is gratefully acknowledged.

**The Conflict of Interest**
The authors declare that there is no conflict of interest regarding the publication of this paper.